# State-of-the-art in selection of variables and functional forms in multivariable analysis – outstanding issues

Willi Sauerbrei[1], Aris Perperoglou[2], Matthias Schmid[3], Michal Abrahamowicz[4], Heiko Becher[5], Harald Binder[1], Daniela Dunkler[6], Frank E. Harrell Jr[7], Patrick Royston[8], and Georg Heinze[6] for TG2 of the STRATOS initiative

[1]Institute of Medical Biometry and Statistics,

Faculty of Medicine and Medical Center - University of Freiburg, Freiburg, Germany

[2]Department of Medical Sciences, University of Essex, Wivenhoe Park, CO43SQ, UK

[3] Department of Medical Biometry, Informatics and Epidemiology, Faculty of Medicine, University of Bonn, Bonn, Germany

[4]McGill University Health Centre, McGill University, Montreal, Canada

[5]Institute for Medical Biometry and Epidemiology, University Medical Center Hamburg-Eppendorf, Hamburg, Germany

[6]Section for Clinical Biometrics, Center for Medical Statistics, Informatics and Intelligent Systems, Medical University of Vienna, Austria

[7]Department of Biostatistics, School of Medicine, Vanderbilt University,

Nashville, Tennesse, US

[8]MRC Clinical Trials Unit at UCL, Institute of Clinical Trials and Methodology, University College London, London, United Kingdom.

**Abstract**

How to select variables and identify functional forms for continuous variables is a key concern when creating a multivariable model. *Ad hoc* 'traditional' approaches to variable selection have been in use for at least 50 years. Similarly, methods for determining functional forms for continuous variables were first suggested many years ago. More recently, many alternative approaches to address these two challenges have been proposed, but knowledge of their properties and meaningful comparisons between them are scarce. To define a state-of-the-art and to provide evidence-supported guidance to researchers who have only a basic level of statistical knowledge  many outstanding issues in multivariable modelling remain. Our main aims are to identify and illustrate such gaps in the literature and present them at a moderate technical level to the wide community of practitioners, researchers and students of statistics.

We briefly discuss general issues in building descriptive regression models, strategies for variable selection, different ways of choosing functional forms for continuous variables, and methods for combining the selection of variables and functions. We discuss two examples, taken from the medical literature, to illustrate problems in the practice of modelling.

Our overview revealed that there is not yet enough evidence on which to base recommendations for the selection of variables and functional forms in multivariable analysis. Such evidence may come from comparisons between alternative methods. In particular, we highlight seven important topics that require further investigation and make suggestions for the direction of further research.

## 1. Introduction

Progress in the theory of mathematical statistics and expansion of computational resources and technology have led to rapid developments in statistical methodology, resulting in many more complex and flexible statistical modelling techniques. Unfortunately, many of these methodological developments are ignored in the analysis of medical data. Consequently, design and analysis of observational studies often exhibit serious weaknesses. To help bridge the gap between recent methodological developments and applications, the STRengthening Analytical Thinking for Observational Studies (STRATOS) initiative was launched in 2013 (Sauerbrei, Abrahamowicz, Altman, le Cessie and Carpenter; 2014). Currently, the initiative has nine topic groups working on various statistical issues in planning and analysis of observational studies. Here we concentrate on issues relevant to topic group 2 (TG2) 'Selection of variables and functional forms in multivariable analysis', which focuses on identifying influential variables and gaining insight into their individual and joint relationships with the outcome.

By definition, work in TG2 is primarily concerned with statistical model building. In this respect, it should be noted that STRATOS also has related topic groups working on prediction (TG6) and modelling issues in causal inference (TG7). In an excellent and thought provoking paper Breiman (2001) differentiated between a data modelling culture and an algorithmic modelling culture. In several comments to this paper (eg Cox (2001), Efron (2001)) differences between possible ways of statistical modelling were discussed. TG2 concentrates mainly on the data modelling culture. As the conflation and confusion between topics and approaches in modelling is common, we will clarify our background and aim by referring to the excellent paper entitled 'To Explain or to Predict' by Shmueli (2010). The latter author nicely explains differences in three conceptual modelling approaches: descriptive, predictive and explanatory modelling. Descriptive modelling is the most commonly used and developed approach, and mainly builds on the data modelling culture. It does not directly aim at obtaining optimal predictive performance, but at capturing the data structure parsimoniously. Nevertheless, a suitable descriptive model is often also an acceptable predictive model. In some fields the term explanatory modelling is used exclusively for testing causal theory. As this paper concentrates on issues in descriptive modelling, aspects of causal modelling will be ignored as causal inferences cannot be drawn from models formulated using a descriptive approach. See Dorie et al (2019) and several comments to this paper for an insightful discussion of the large number of issues in causal inference. We note that in the context of descriptive modelling some of us used terms like 'explanation' in earlier papers and a book (Sauerbrei, Royston, Binder, 2007; Royston and Sauerbrei, 2008). For a more detailed discussion of differences between the three types of modelling, please see Shmueli (2010).

The intended use of a model is a key consideration when choosing a suitable analysis strategy. For a descriptive model, interpretability, transportability and general usability are important criteria. By contrast, if descriptiveness plays no role, models for prediction may be optimized for predictive accuracy and can then be more complex (more variables included, more complex functions allowed). With very complex and interrelated associations of covariates with the outcome and enough data to unravel such associations, algorithmic modelling approaches may become valuable modelling tools. No matter if descriptiveness is important or not, biases are introduced by deriving the 'final' model data-dependently. However, the relevance of critical issues (such as biases, large variances, number of variables included and model complexity) must be weighed against advantages with respect to the intended purpose of modelling.

To obtain good descriptive models, the (interrelated) challenges of selection of variables for inclusion and choice of the functional forms for continuous variables must be tackled (Harrell, 2015; Royston et al, 2008), and this constitutes the main aim of TG2. In general, the analyst is expected to conduct the analysis according to 'state-of-the-art (SOTA)' methodology, which is defined in Wikipedia (Wikimedia, 2019) as 'State-of-the-art refers to the highest level of general development, as of a device, technique, or scientific field achieved at a particular time.' SOTA also plays a key role in applications for funding of a research project, where a brief section summarizing SOTA is often requested. However, do we have enough evidence to define SOTA methodology for the selection of variables and functional forms for continuous variables?

About three decades ago Picard and Cook (Picard and Cook, 1984) stated 'Exact distributional results [on maximum likelihood estimates and test statistics] are virtually impossible to obtain [because of data-dependent model building], even for the simplest of common subset selection algorithms'. Although there is substantial progress, current developments and empirical comparisons are inadequate to identify advantages and disadvantages of the many variable selection techniques and to provide evidence-supported arguments regarding SOTA strategies for variable selection.

For more than 50 years suitable algorithms have been available for various stepwise variable selection techniques (Efroymson, 1960). Triggered by the intention to derive suitable 'omics' predictors (often called 'gene signatures') in high-dimensional data with many more explanatory variables than observations, various techniques for selecting variables have been proposed during the last two decades, adding to the long list of 'traditional' approaches. Consequently, analysts may have access to many toolkits, but it is not straightforward which variable selection approach they should use and under what circumstances. Will the results be (seriously) influenced by the variable selection method chosen?

Similar arguments apply to the methodology for determining the functional form for continuous variables such as age, blood pressure, values of laboratory tests, exposure intensity or dose of a substance which often play an important predictive or explanatory role. Assuming linear effects of such variables on the outcome is often the default choice and rarely questioned, but the reason for this is often ignorance and not the result of carefully considering the most plausible functional form. Hence, the variable selection problem needs to be combined with the selection of suitable functional form(s) for continuous variables, i.e. the dose-response function, substantially increasing the complexity of the modelling issues. For function selection, at least four general approaches, with several variations, are used in recent studies: assuming *a priori* a linear relationship; categorization; fractional polynomials; and different spline-based methods. Yet, while all rely on important explicit or implicit assumptions and have some limitations, there is little guidance to choose between these approaches. Reasons for selecting one or the other are rarely explained. Several methods have been published, but the number of studies comparing alternative strategies is limited and guidance is insufficient. Consequently, there are strong barriers concerning knowledge transfer (Pullenayegum et al, 2015).

With a focus on applications in observational studies of human health, we provide a brief overview of existing methods for variable selection, identify some issues with each of the four approaches to represent the functional form for continuous variables, and discuss topics requiring further research. The goal is to provide evidence-supported arguments regarding SOTA methodology for building multivariable models with continuous variables. Experienced statisticians belonging to the STRATOS 'level 2' category (Sauerbrei et al, 2014; Table 1) are the intended audience for this paper.

In section 2, we discuss general issues in building descriptive regression models. In section 3, we briefly summarize most popular methods of variable selection. Approaches to handle continuous variables are outlined in section 4, and the combination of methods for variable and functional form selection is discussed in section 5. In section 6, we provide two examples of potentially suboptimal approaches to illustrate weaknesses in analyses related to TG2 issues. In section 7, we identify some unsolved methodological issues that require further research towards establishing SOTA methodology, before giving concluding remarks in section 8.

## 2. General issues in building descriptive regression models

In most observational studies, a mix of continuous and categorical variables are collected. Many of these variables are correlated, implying that the results of selection and/or modelling of a given variable may depend, sometimes critically, on whether and how some of the other variables are represented in the multivariable model. The current paper focuses on popular multivariable regression models, such as a normal-errors regression model for continuous outcomes, a logistic regression model for binary outcomes, or the Cox proportional hazards model for censored survival data. Arguably, these are the main types of regression models used in health sciences applications. In health research, depending on the context, the terms risk factor, prognostic factor, predictor, exposure, covariate, or confounder are often used for variables. Here we will use them interchangeably.

It is generally acknowledged that subject-matter knowledge should guide multivariable model-building. A study should be carefully planned, guided by the research questions and consideration of the methods envisaged for analysing the data. In randomized trials a detailed analysis plan is written before the final data are collected and the analysis starts. However, the situation is more complex with observational studies (Royston et al, 2008). For example, when considering the association between a continuous exposure and the probability of developing a disease, subject-matter knowledge does not always give clear advice for the set of potential confounders that the model should be adjusted for, nor in particular for the functional forms of their association with the outcome (Benedetti and Abrahamowicz, 2004). Since subject-matter knowledge is often limited or at best fragile, data-dependent model building is necessary (Harrell, 2001). Pre-specifying all aspects of a model is almost invariably unrealistic in observational studies. The gain of preventing bias by avoiding data-dependent model building is offset by the high cost of ignoring models which may fit the data much better than the pre-specified one. Also, what if assumptions underlying the pre-specified model are violated?

A key issue of variable selection is which of the candidate variables to include in the 'final' model. Some researchers argue that variables with 'strong' effects will be included by any sensible variable selection strategy and that it is less important to include a larger number of variables with no or a 'small' effect (Sauerbrei, Buchholz, Boulesteix and Binder, 2015). The potential cost of obtaining information on all selected variables also impinges on the general usability of a model. If the originally selected model indicates that a variable whose values are difficult to obtain has only a small effect on the outcome, such a model will be infrequently used in practice.

Breiman (1992) states 'When a regression problem contains many predictor variables, it is rarely wise to try to fit the data by means of a least squares regression on all of the predictor variables. Usually, a regression equation based on a few variables will be more accurate and certainly simpler.'

Variable selection methods aim to eliminate all uninfluential variables, at least at a defined evidential level (e.g. a nominal p-value). Data-dependent model building certainly introduces several types of bias and problems (Steiner and Kim, 2016; Heinze, Wallisch and Dunkler,

2018), including replication stability, selection bias, omitted variable bias, underestimation of variance and overestimation of the predictive ability of a model. These are critically important for weak risk factors, but 'weak' must be judged in the context of the sample size. The possible magnitude of these biases must be weighed against the advantages of better interpretability of smaller models.

Another crucial issue for model building is the functional form with which variables should enter the variable selection procedure. In practice, continuous variables are often categorized before being subjected to variable selection, although major weaknesses of categorizing continuous variables have been well-known for decades (Altman, Lausen, Sauerbrei and Schumacher, 1994; Royston, Altman and Sauerbrei, 2006). Aside from categorisation, assuming linearity is the norm. Flexible modelling techniques based on various types of smoothers, including fractional polynomials and several types of splines, are available but are still underused. Although selection of variables and functional forms for continuous variables both apply to many analyses, suitable methods that address both issues are not often utilized.

Of course, many more analysis issues arise, such as the investigation of potential interactions between two or more predictors or interactions between a predictor and follow-up time in a survival analysis (time-dependent effects). We do not consider them in what follows. Finally, we stress that many decisions are required before model building can start. For a discussion of several issues we refer to Chatfield (2002), to the paper of Mallows (1998) on the 'zeroth' problem, and to the first paper by the STRATOS topic group TG3 (Initial data analysis), which discusses a conceptual framework to handle the large number of issues (Huebner, le Cessie, Schmidt and Vach, 2018).

### 3. Variable selection: strategies

Virtually any statistical software package contains procedures for variable selection. This made their use very popular, particularly with end-users without a formal background in statistics. However, this wide availability has been a breeding ground for common misapprehensions of the role and necessity for variable selection (Heinze et al, 2018).

### 3.1 Traditional variable selection strategies

Traditional variable selection methods involve automated iterative stepwise model building procedures that rely on testing the statistical significance of the regression coefficients for all variables considered at a given step. One popular approach starts from a null model containing no variables, and adds one variable at a time, corresponding to the most significant (lowest p-value) of their contributions to improving model fit (forward selection, FS). Alternatively, a 'full model' including all potential explanatory variables is estimated first and then the least significant variables are eliminated one-by-one (backward elimination, BE). The 'stepwise' approach implemented in some software packages combines forward selection with additional backward steps. The significance level used as the stopping threshold is the tuning parameter that determines the size of the final model (Royston et al, 2008). The smaller the level, the fewer variables will be selected.

Biostatisticians widely agree that among the selection procedures just described, backward elimination, with or without additional forward steps, is the preferable method as it already starts with a plausible model (Mantel, 1970). The backward elimination algorithm sets some regression coefficients exactly to zero contrary to the actual (unpenalized) maximum likelihood estimates of the regression coefficients (which are usually never exactly zero). This means that one intentionally moves away from the maximum likelihood solution by inducing bias. Particularly for weak predictors, this bias may decrease the variance because of an implicit shrinkage towards zero. However, if a weak predictor is 'selected' by the algorithm, its corresponding regression coefficient may be seriously biased away from zero (Miller, 1984; Copas and Long, 1991; Sauerbrei, 1999). The increased bias usually also increases the mean squared error (MSE) of the predicted value of the outcome, and the random element contributed by an uncertain selection decision further contributes to MSE inflation. In a recent review, Heinze et al (2018) proposed some resampling-based measures to estimate the bias conditional on selection and the inflation in (root) MSE caused by the selection. Although intuitively appealing, their proposals are still to be validated in simulation studies.

Backward elimination and other stepwise selection methods are shortcuts to the computationally more expensive best subset selection, where all possible models are estimated and their fit is compared using some information criterion. Different information criteria have been proposed, among which Akaike's information criterion (AIC) is in principle preferable for predictive models, and Schwartz's Bayesian information criterion (BIC) for descriptive models (Burnham and Anderson, 2002). The purpose of AIC is to approximate the predictive performance, quantified by the model deviance, which would be expected in validation in an independent sample drawn from the same population. By contrast, BIC aims at selecting the true predictor variables (Raftery, 1995). However, BIC cannot reach this goal unless the model space searched includes the 'true data generating model' and the data set is large enough for such an approximation to work. For comparison of two hierarchically nested models differing in only one model term with one degree of freedom (df), AIC and BIC can be directly translated into significance tests of that model term. For AIC, the corresponding threshold p-value is 0.157. For BIC the threshold depends on the sample size; the penalty parameter is $\log(n)$ (Teräsvirta and Mellin, 1986). For example, sample sizes of 100 or 400 correspond to significance levels of 0.032 or 0.014. Thus, while AIC will include more and more (weak) predictors with increasing sample size, selection by BIC will become more and more strict. Originally these criteria were developed to compare a small number of pre-specified models, not to screen a large model space. However, nowadays they are used more generally as a pragmatic approach to model selection.

Best subset selection may not pose a big computational challenge in the era of parallel computing, but even so, it may not be preferable to backward elimination. The increased flexibility may engender a higher probability of selecting 'spurious' predictors, i.e., to increase selection uncertainty. Furthermore, if variable selection is to be combined with selection of the functional forms of continuous variables (see section 5), a best subset selection procedure would increase the number of candidate models substantially. To the best of our knowledge, it has never been suggested.

'Univariate selection', i.e. selection of variables based on significance in univariable regression models, is still popular, notably in high-dimensional studies in the fields of bioinformatics and (epi-)genetics (Winter et al, 2012; Fröhlich, 2014). For example, the FDA-recommended CKD273 signature for diagnosis of early kidney disease is based on univariate protein filtering followed by a machine-learning algorithm applied to the filtered features (Good et al, 2010; Dakna et al, 2010; Nkuipou-Kenfack, Zürbig and Mischak, 2017). Among non-statisticians, it is still even believed that univariate significance is a prerequisite for including a variable in a multivariable model (Heinze and Dunkler, 2017). In contrast, it is generally accepted among statisticians that the difference between unadjusted and adjusted effects of a variable can go in either direction, therefore that univariable selection may be misleading (Sun, Shook and Kay, 1996).

### 3.2 Procedures based on 'change-in-estimate'

Selection procedures based on the change-in-estimate are particularly popular in epidemiology, especially for the situation where the effect of an exposure of main interest should be adjusted for confounders (Maldonado and Greenland, 1993; Lee, 2014; Groenwold, Klungel, van der Graaf, Hoes and Moons, 2013). The change-in-estimate criterion evaluates if elimination of a potential confounder from the model would lead to a relevant change in the regression coefficients of the exposure variable of interest. 'Relevance' is often defined as more than, e.g., a 10% change in the regression coefficient of the exposure of interest (Lee, 2014). Hosmer, Lemeshow and colleagues incorporated the change-in-estimate criterion into significance-based selection procedures, yielding their 'purposeful selection' (Hosmer, Lemeshow and May, 2008; Hosmer, Lemeshow and Sturdivant, 2013; Bursac, Gauss, Williams and Hosmer, 2008). This was later revisited by

Dunkler et al (Dunkler, Plischke, Leffondré and Heinze, 2014), who proposed modifying the change-in-estimate. In their 'augmented backward elimination' procedure. Non-significant variables are not omitted from a model if a standardized change in estimate following their elimination exceeds a certain pre-specified threshold. They argue that their algorithm will generate final models that are close to the starting model and will only eliminate really unnecessary variables. Therefore, the resulting bias is likely to be smaller than that induced by the conventional backward elimination procedure.

### 3.3 Modern variable selection strategies

#### 3.3.1 Penalized likelihood

Two approaches to combining variable selection with shrinkage were proposed in the mid-1990s. The non-negative garrotte starts with ordinary least squares estimates from the full model and estimates parameterwise shrinkage factors under a specified constraint (Breiman, 1995). For coefficients with a stronger effect, the shrinkage factor will be close to 1. For variables that are weakly associated with the response, the least-squares estimate is likely to be small and consequently the shrinkage factor will tend towards zero, corresponding to variable elimination. As an estimate from the full model is needed, the approach cannot be used for the analysis of high-dimensional data. We consider this as one of the main reasons that the approach did not become more popular.

Shortly after the non-negative garrotte, Tibshirani (1996) proposed the popular Lasso method, which consists of penalizing the log likelihood of a model with a multiple of the sum of absolute regression coefficients. Consequently, some of the regression coefficients become exactly zero and hence the method provides variable selection. The strength of the penalty, i.e., the multiplier of the penalty term, often symbolized by $\lambda$, determines the size of the resulting model. Usually $\lambda$ is estimated from the data. It can be determined in various ways, cross-validation being most common. A practical problem with the Lasso penalty is that the regression coefficients of different variables should all have the same scaling. This is usually achieved by working with standardized covariates, i.e., by first dividing each covariate by its empirical standard deviation. However, different choices for standardization may lead to different results. Therefore, one should bear in mind that the Lasso, unlike the traditional significance-based selection procedures, is not invariant to transformations of the covariates. Nowadays the Lasso is widely accessible because of many implementations in R (the `glmnet` package being the most popular (Friedman, Hastie and Tibshirani, 2010), and in other software packages.

The Lasso tends to screen variables rather than to select them, meaning that it often includes variables with very weak effects. Furthermore, in the presence of collinear or grouped variables, the Lasso fails to select all variables in the group. For high-dimensional data where 'p>n', the Lasso can only select up to n variables. To address these issues, modifications have been suggested including the Elastic Net (Zou and Hastie, 2005) and the Adaptive Lasso (Zou, 2005). The Elastic Net combines L1 (equivalent to the Lasso) and L2 (Ridge) penalization, removing the limitation on the number of selected variables but also encouraging grouped selection. An advantage of the Elastic Net is that it includes L1 and L2 penalization as special cases. The Adaptive Lasso penalizes the log likelihood with $\lambda$ times a weighted sum of absolute regression coefficients, where the weights are determined in a pre-estimation step, usually as the reciprocal absolute regression coefficients from a maximum likelihood or ridge regression fit. The rationale for this is to down-weight penalties of strong predictors and to up-weight weak predictors in the penalty term. Usually it results in models with fewer variables and stronger effects.

Another popular modification with the same goal is Smoothly Clipped Absolute Deviation (SCAD; Fan and Li, 2001), where coefficients larger in an absolute sense are penalized less than smaller ones. Further generalizations of the Lasso are listed in Tibshirani (2011) and Hastie et al (Hastie, Tibshirani and Wainwright, 2015).

### 3.3.2 Boosting

Component-wise boosting is a forward stagewise regression procedure applicable to generalized linear models, Cox models, quantile regression, and many other types of statistical regression models (Tutz and Binder, 2006; Bühlmann and Hothorn, 2007; Binder and Schumacher, 2008). The method is rooted in the field of machine learning, where it was first used for tree-based binary classification (AdaBoost algorithm, Freund and Schapire, 1996). Based on work by Friedman et al. (Friedman, Hastie and Tibshirani, 2000), who discovered the connection between boosting and logistic regression, many types of boosting algorithms have been proposed. Today, two popular approaches are gradient boosting (Friedman, 2001; Bühlmann et al, 2007) and likelihood-based boosting (Tutz et al, 2006), which coincide in the case of Gaussian regression but differ in their applicability and assumptions otherwise (see Mayr, Binder, Gefeller and Schmid, 2014 for a comparison). Regardless of their conceptual differences, both approaches make use of the same basic idea for variable selection: Starting with an empty model, only one regression coefficient is updated in each step of the boosting algorithm ("component-wise" updates). In contrast to traditional FS, updates are done by adding only small fractions (e.g., one tenth or one hundredth) of the estimated regression coefficients to the estimates obtained in previous steps ("stagewise" procedure). Variable selection (i.e., selection of the regression coefficient to update) is done by evaluating in each step all univariable regression models conditional on the estimates of the previous steps (used as offsets) and to select the variable corresponding to the best-fitting univariable model. Stopping the boosting algorithm before convergence ("early stopping") results in variable selection and shrinkage, like the Lasso. In fact, boosting and the LASSO are equivalent in specific settings and can even be shown to be modifications of the same underlying algorithm (least angle regression (LARS), Efron, Hastie, Johnstone and Tibshirani, 2004), which explains their similar performance in many applications. The number of boosting steps is the main tuning parameter of the procedure, often optimized by cross-validation. Boosting algorithms are particularly useful in high-dimensional settings but can also be applied in the low-dimensional case.

### 3.4 Resampling-based variable selection procedures

Since the mid-eighties of the last century, resampling procedures (usually the nonparametric bootstrap) have been proposed to assess model (in-)stability. The bootstrap inclusion frequencies (BIF) from a large number of resampling replications is used as a criterion for model selection (Gong, 1982; Chen and George, 1985; Altman and Andersen, 1989). These approaches are based on the idea that variables with a "stronger effect" are selected with a high probability in each replication, whereas the selection of others may be a matter of chance. A BIF threshold needs to be chosen and only variables with larger BIFs are included in the model.

However, this simple 'summary' of the selected models from resampling replications may result in selecting a 'bad' model if the inclusion of two (or more) variables depends on each other. One important reason is the correlation structure of the candidate variables, which may strongly influence BIFs of some variables. It may happen that one of two variables is included in each of the replications, but the individual BIFs are not much higher than 50%. Thus, based on its BIF, neither variable may be judged important. To cope with such situations, Sauerbrei and Schumacher (1992) suggested considering dependencies among inclusion fractions. They defined two bootstrap variable selection procedures, one aiming for a simpler model including strong factors only, whereas the other including also weaker factors. The BIFs and their dependencies are used as criteria to select a final model. Recently, a large simulation study showed promising results (De Bin and Sauerbrei, 2018), but more practical experience is needed. However, the issue of dependencies among variable inclusions seems to be widely ignored and further issues need more consideration (such as the central role of the originally used variable selection procedure and the effect of influential points).

In the last decade, resampling-based variable selection procedures have become more popular. Meinshausen and Bühlmann (2010) introduced stability selection, which combines subsampling (bootstrap without replacement) with (high dimensional) selection algorithms. The method is very general and is widely applicable. The authors state that stability selection provides finite sample control for some error rates of false discoveries. Recently, the specific type of resampling procedure has become a topic of research. Some argue that subsampling may be preferable to the nonparametric bootstrap for investigations in the context of model building (De Bin, Janitza, Sauerbrei and Boulesteix, 2016; Janitza, Binder and Boulesteix, 2016; Rospleszcz, Janitza and Boulesteix, 2016). However, issues like the most suitable sampling rate (0.632 seems to be the most common) need further investigation.

### 3.5 Bias and the role of shrinkage methods

The regularized procedures described in Section 3.2 are so-called shrinkage methods, as they introduce some bias towards zero into the regression coefficients to try to reduce the prediction error. Shrinkage methods can also be applied in combination with traditional variable selection methods, where the use of shrinkage factors has been proposed to improve the accuracy of predictions. In the latter context, shrinkage factors may be estimated using resampling and are then multiplied by the regression coefficients to obtain shrunken coefficients. Either a common global shrinkage factor (Van Houwelingen and le Cessie, 1990) or parameter-specific shrinkage factors (Sauerbrei, 1999) can be used for this purpose. Van Houwelingen and Sauerbrei (2013) revisited these approaches in the context of linear regression models and found that backward elimination in combination with parameter-specific shrinkage factors yields models that are similarly or even more accurate than those from the modern approaches. These shrinkage methods may also act as a tool to (partly) remove the overestimation bias induced by variable selection methods (discussed in Section 3.1), even though their construction is not motivated by the goal of bias correction. A systematic investigation of this 'side effect' is still lacking. Moreover, the proposed estimation method by leave-one-out cross-validation could be questioned, as in other contexts, other resampling procedures have been preferred (Smith, Seaman, Wood, Royston and White, 2014).

### 3.6 Post-selection inference

Since the main aim of descriptive models is the interpretation of the estimated regression coefficients, point estimates should be accompanied by confidence intervals and p-values. For traditional selection procedures, virtually any statistical software reports confidence intervals and p-values of the final model, derived under the assumption that the model was given *a priori*. This ignores uncertainty in model predictions, estimates of effects, and variance caused by model selection. Breiman (1992) called this a "quiet scandal". Meanwhile, underestimation of the uncertainty has been confirmed in numerous simulation studies for different types of models. Various approaches for coping with model uncertainty have been proposed, starting with Chatfield (1995) and Draper (1995). Recently, Heinze et al (2018) proposed for different settings: 1) Use the full model for inference about regression coefficients, where the full model is the model including all variables considered for adjustment, and so each regression coefficient is adjusted for all others. 2) If a small number of pre-specified competing models is evaluated, use model-averaging as proposed by Burnham and Anderson (2002). 3) If the full model is implausible and no evidence for a small set of candidate variables is available, assess the variability of the regression coefficients around the full model estimates using the bootstrap, where in each bootstrap replication the selection procedure is repeated. When comparing this variability with the standard error in the full model, one often sees variance inflation, even for strong predictors, and variance reduction only for highly non-influential variables (Heinze et al, 2018).

Although statistical inference after data-driven model selection has been deemed impossible (Leeb and Pötscher, 2005), recent years have seen a couple of advances in that direction (Taylor and Tibshirani, 2015). There are now even some R packages available with which confidence intervals for regression parameters selected with the Lasso can be estimated (Tibshirani, Taylor, Loftus and Reid, 2017). However, given that these procedures were developed only recently, evidence of their performance in practice is still limited.

Parameterwise or joint shrinkage factors (Sauerbrei, 1999; Dunkler, Sauerbrei and Heinze, 2016) have been proposed in an attempt to reduce overestimation bias due to selection of variables and functional forms. It is still unclear how a correction for the variance, e.g., by using a sandwich variance estimator that is robust to model misspecification (White, 1980a; White, 1980b), may help in achieving approximately valid confidence intervals.

## 4. Continuous variables – to categorize or to model?

The effects of continuous predictors are typically modelled by either categorizing them (which raises issues such as the number of categories, cutpoint values, implausibility of the resulting step-function relationships, local biases, power loss, or invalidity of inference due to data-dependent cutpoints) (Greenland, 1995) or assuming linear relationships with the outcome, possibly after a simple transformation (e.g. logarithmic or quadratic). Often, however, the reasons for choosing such conventional representations of continuous variables are not discussed and the validity of the underlying assumptions is not assessed (Sauerbrei et al, 2014). Commonly used methods are summarized by Becher (Becher, 2014).

To address limitations caused by categorization or a questionable assumption of linearity, statisticians have developed flexible modelling techniques based on various types of smoothers, including fractional polynomials (Royston and Altman, 1994; Royston et al, 2008) and several types of splines. The latter include restricted regression splines (Harrell, 2015; de Boor, 2001), penalized regression splines (Wood, 2006) and smoothing splines. For multivariable analysis, these smoothers have been incorporated in generalized additive models (Hastie and Tibshirani, 1990). Depending on personal preferences of the analyst, usually one approach for the main analysis is chosen, sometimes complemented by alternative approaches as sensitivity analyses. Some properties of the approaches, which may lead to different results, are given below.

## 4.1 Categorization

Categorization allows researchers to avoid strong assumptions about the functional relationship between the continuous variable and the outcome of interest. Some analysts incorrectly consider categorization to be 'model-free'. It was often the method of choice in the past when analysis was based on contingency tables. The approach has the advantage that it leads to results that seem to be easier to interpret and is therefore favoured by non-statisticians. The functional form for the effect of a covariate is often unknown, and an analysis using a categorized variable is simpler than working with continuous variables directly. In general, reporting is easier, and categorization allows graphical presentation, e.g.

Kaplan-Meier curves can be produced in survival analysis. However, it has some serious drawbacks.

Categorization raises several issues such as the number of cutpoints, where to place them and how to code the resulting variable for the analysis. Obviously, use of two groups makes it impossible to detect a non-linear relationship. It is well-known that the results of analyses can vary if different cutpoints are used for splitting. Usually dummy variables are defined and different codings are possible. Alternatively, a coding scheme with a suitable choice of scores implying a metric (e.g. {1, 2, 3}, {1, 2, 4} or scores representing the median of each category), together with a test for linear trend, is possible. For variables with a spike at zero (see section 4.3) a natural baseline category is the unexposed group.

Many objections to categorization have been raised during the last three decades. It is often stressed that categorization should not be used to derive the final model (Miller and Siegmund, 1982; Hilsenbeck, Clark and McGuire, 1992; Altman et al, 1994; Royston et al, 2006; van Walraven and Hart, 2008; Vickers and Lilja, 2009). The full information contained in the data is not used and the resulting risk function is a step function. From a biological perspective, a step function may be unrealistic, since individuals close to but on opposite sides of a cutpoint are characterized as having very different risks or prognostic scores. If the (unknown) underlying relationship with the outcome is expected to be smooth but not necessarily linear, interpretation of step functions is implausible. Furthermore, categorization discards information, the greatest loss occurring when a variable is dichotomized. However, when interest centres on accurate prediction at small or large values of a variable, van Houwelingen (2014) produced arguments favouring grouping.

Some analysts appear to believe that the severe information loss incurred by categorization may be compensated by finding the 'optimal' cutpoint that defines two groups. Every possible cutpoint on x is considered and the value of x which minimizes the p-value is chosen. The cutpoint selected is to some extent due to chance and is not reproducible. Altman et al. (1994) called this procedure the 'minimum p-value' approach and stressed that the estimated effect describing the difference between two groups is strongly overestimated and that if uncorrected, multiple testing increases the type I error to around 40% instead of a nominal 5%.

Other methods of determining a cutpoint are to use recognized values (e.g., > 25 kg/m$^2$ to define 'overweight' based on body mass index), to use 'round numbers' such as multiples of five or ten, to use the upper limit of a reference interval in healthy individuals, or to apply cutpoints used in previous studies. These methods still incur information loss, but at least not an inflated type I error probability. In the absence of predefined cutpoints, epidemiological researchers often use three to five categories based on quantiles of the empirical distribution, while in clinical studies the most common choice is dichotomization, often at the sample median. However, different cutpoints are used in different studies, so the results can neither be easily compared nor summarized. For example, nineteen cutpoints were identified when the prognostic value of S-Phase fraction in breast cancer patients was assessed (Altman et al, 1994). Several of them resulted from an 'optimal' cutpoint analysis.

Categorizing a continuous variable into three or more groups reduces the loss of information. Several methods of coding can be used for the categorized variable. Care is needed to choose the coding in a suitable way, specifically if variable selection is used in combination with dummy variables (4, section 3.3). Despite the well-known problems caused by categorization, it is still the preferred approach in practice. Descriptive analysis of continuous covariables is often done with frequency tables for which some form of grouping is needed. In a recent review of cohort studies in physical activity, the primary exposure was categorized in 21/30 (70%) of studies. In 51 dietary intake studies, all variables were categorized in 27 (53%) and at least one variable in 50 (98%) (Shaw et al, 2018).

To summarize, categorization of a continuous or semi-continuous variable may be useful for descriptive presentation of data and for initial analysis. Furthermore, categorization is often required in clinical decision making. However, it is unnecessary and often harmful when used to build the final statistical model. Several modelling approaches have been proposed and should become the standard to assess the functional influence of a continuous variable, avoiding the need to categorize.

## 4.2 Modelling nonlinear effects of continuous variables

In the following we will briefly describe and discuss the fractional polynomial and spline-based approaches. For more details we refer to the web supplement and the literature referenced in the following paragraphs.

### 4.2.1 Fractional polynomials

The class of fractional polynomial (FP) functions is an extension of power transformations of a covariate. For most applications FP1 ($\beta_1 x^{p1}$) and FP2 ($\beta_1 x^{p1} + \beta_2 x^{p2}$) functions are sufficient.

For the exponents $p_1$ and $p_2$ a set S={-2, -1, -0.5, 0, 0.5, 1, 2, 3}, with 0 = log x has been proposed. For $p_1 = p_2 = p$ ('repeated powers') an FP2 function is defined as $\beta_1 x^p + \beta_2 x^p \log x$. This defines 8 FP1 and 36 FP2 models. The class of FP functions appears small, but it includes very different types of shapes (Royston et al, 1994). General FPm functions are well-defined and straightforward, but are rarely used in medical applications. The main aim of FP modelling is the derivation of a suitable function which fits the data well, while being simple, interpretable and generally usable. Therefore, a linear function is the default and a more complex non-linear function is only chosen if strongly supported (according to a chosen significance level) by the data. The functional relationship is selected by a function selection procedure (FSP) with up to three steps (if FP2 is the most complex function considered) (Royston et al, 2008). The test at step 1 is of overall association of the variable with the outcome. The test at step 2 examines the evidence for nonlinearity. The test at step 3 chooses between a simpler FP1 or a more complex FP2 nonlinear model (for details see the web supplement).

In section 2 we emphasized the need for a model building procedure related to the main aim of the analysis. For FPs this can be done by pre-specifying the most complex FP functions allowed (often FP2, sometimes FP1, occasionally FPm with m > 2) and by choosing nominal significance levels, which are the key tuning parameters for determining the function's complexity according to the aim of the study.

It is well-known that data-dependent model building results in biased estimates and incorrect p-values. However, the FSP uses the principle of a closed test procedure which ensures that the overall type 1 error is close to the nominal significance level (Marcus, Peritz and Gabriel, 1976). For some simulation results on type 1 error and power, see Royston et al(2008), section 4.10.5.

## 4.2.2 Splines

Another common method of modelling nonlinear effects of continuous variables is spline regression. Generally, a spline function is defined by a set of piecewise polynomial functions of a continuous variable that are joined smoothly at a set of knots spread across the support of the variable (Wood, 2017 In contrast to fractional polynomials, which are constructed by weighted sums of power functions with exponents -2, -1, -0.5, 0, 0.5, 1, 2, 3, the piecewise polynomial functions defining a spline all have the same degree. For example, *cubic* splines are constructed from a set of piecewise polynomials each of degree 3. Another conceptual difference between splines and fractional polynomials is the range over which the components of the two methods are defined. Whereas splines are constructed from 'local' piecewise functions (each defined between two consecutive knots), the power functions involved in the construction of a fractional polynomial are each defined over the whole range of the continuous variable.

Generalized additive models (GAMs, Hastie and Tibshirani, 1990; Wood, 2017). GAMs extend generalized linear models by the inclusion of smooth nonlinear effects of continuous variables. Each nonlinear effect is represented by a spline function (although other smoothers are in principle possible). Estimation of the spline functions is usually carried out by writing splines as a linear combination of spline basis functions and by estimating the weights in this combination via (possibly penalized) maximum likelihood optimization.

Generally, GAMs support a huge number of modelling alternatives and model fitting techniques. This is because splines come in many shapes and forms differing, e.g., in the degree of the polynomial pieces and/or the number and positions of the knots. Furthermore, it is common to impose restrictions on the spline functions, for example linearity restrictions at the boundaries of the variables' supports (used e.g. in natural splines, Hastie, Tibshirani and Friedman, 2009), complexity restrictions imposed by lower-order approximations (e.g. in thin-plate regression splines; Wood, 2003), smoothness restrictions imposed by 'wiggliness' penalties (used e.g. in smoothing splines and in P-splines, which penalize the second- or higher-order differences of the basis function coefficients, Eilers and Marx, 1996), or imposing monotonicity (Ramsay, 1988). If penalties are used in GAMs, each spline function is usually associated with a separate hyperparameter determining the weight of the respective penalty in penalized maximum likelihood estimation. In addition, the definition of the basis functions is not unique, and there exist various possibilities on how to define them, each with different numerical properties. For details, see the web supplement. Note that splines can also be used in boosting algorithms, see e.g. Schmid and Hothorn (2008). Obviously, all the aforementioned modelling options affect the shapes of the spline functions, and hence also the effects and the interpretation of the respective variables in a GAM. Perperoglou et al (Perperoglou, Sauerbrei, Abrahamowicz and Schmid, 2019) provide an overview of the most widely used spline-based techniques and their implementation in R.

In an investigation of the effect of alcohol intake as a risk factor for primary oral cancer, Rosenberg et al (Rosenberg et al, 2003) discussed several issues of spline modelling when determining the functional form of a continuous variable in practice. Abrahamowicz et al (Abrahamowicz, du Berger and Grover, 1997) illustrated the practical advantages of spline modelling, implemented with the smoothing splines within GAMs, to detect and account for nonlinear associations in the context of cardiovascular epidemiology. They demonstrated how the flexibility of the splines may enhance the robustness of the estimates.

### 4.3 FP procedure to handle variables with a 'spike at zero'

In epidemiology and in clinical research, variables often have unusual distributions. Commonly in epidemiological studies, a proportion of individuals have zero exposure to the risk factor of interest, whereas for the remainder, the distribution of the risk factor is continuous. We call this a 'spike at zero' or a 'semi-continuous variable'. Examples are smoking, duration of breastfeeding, or alcohol consumption. Furthermore, the empirical distribution of laboratory values and other measurements may have a semi-continuous distribution because the lower detection limit of the measurement is assigned a value of zero. Such a variable cannot directly be analysed with FP's (see 4.2.1) since some of the transformations needed are only defined for positive values. An *ad hoc* but arguably unsatisfactory method is to add a small constant to each observation.

Spike at zero variables can be analysed by categorization, using the zero category as baseline. The method has the same disadvantages as mentioned in section 4.1. A better approach is described (Becher, Lorenz, Royston and Sauerbrei, 2012; Becher, 2014) and illustrated (Lorenz, Jenkner, Sauerbrei and Becher, 2017), using an extended version of fractional polynomials. Briefly, a binary indicator variable Z (exposed/non-exposed) is generated, and the modelling involves both Z and the spike at zero variable X, where the transformations are applied to the positive part of X only. A similar approach was proposed earlier in the context of parametric modelling of the association between smoking intensity and lung cancer. Leffondre et al (Leffondre, Abrahamowicz, Siemiatycki and Rachet, 2002) demonstrated that it allows researchers to separate (i) the 'qualitative' difference between ever-smokers relative to non-smokers, from (ii) the 'quantitative' dose-response association estimated using data on smokers only.

Four closely related procedures have been proposed for model building with two (which can be extended to more) spike-at-zero variables (Jenkner, Lorenz, Becher and Sauerbrei, 2016). Although it is relevant to the analysis of many studies, practical experience is limited.

## 5. Combining variable and function selection

Variable selection in the presence of non-linear relationships of covariates with the outcome is an even more complicated exercise. In fact, decisions regarding the inclusion/exclusion of specific variables and modelling of the functional forms of both these variables and potential confounders may depend on each other in a complex way. Firstly, the 'importance' and/or statistical significance of a continuous variable may depend strongly on how its relationship with the outcome is modelled. For example, whereas there was no evidence for an association of body mass index (BMI) with the logit of the probability of coronary heart death (CHD) assuming a linear effect, a GAM-based multivariable model with the same adjustment covariates revealed a non-monotone J-shaped association. CHD mortality increased for both very lean and obese subjects (Abrahamowicz et al, 1997).

## 5.1 The multivariable fractional polynomial approach

The multivariable fractional polynomial (MFP) approach is a pragmatic procedure to create a multivariable model with the parallel aims of selecting important variables and determining a suitable functional form for continuous predictors (Sauerbrei and Royston, 1999; Royston et al, 2008). To select a model, significance levels chosen for the two components (backward elimination (BE) and FP functions) are the key settings for determining model complexity. If more or fewer variables are included, simple linear or more complex non-linear functions are selected. If an extremely low significance level (e.g. 0.00001) is used to select FP functions, usually only linear functions are selected and MFP in effect becomes just BE. Choosing 0.999 as a significance level in the BE component ensures that a specific variable is guaranteed to enter the model. Although relatively simple and easily understood by researchers familiar with the basics of regression models, the selected models often extract most of the important information from the data. Models derived are relatively easy to interpret and to report, a pre-requisite for transportability and general use in practice. Easy to use software in Stata, SAS and R is available (Sauerbrei, Holländer, Riley and Altman, 2006). Several extensions of MFP have been proposed (e.g. MFPI for interactions in RCTs, MFPT to model time-dependent effects in survival data, metacurve for meta-analysis of functions). Only programs in Stata are available for these extensions. For more details see http://mfp.imbi.uni-freiburg.de/.

Because of data-dependent selection of variables and their functional forms, overestimation bias is likely to arise. As with the arguments in the context of variable selection (section 3.4), bias may be reduced by estimating and applying parameter-specific shrinkage factors (see section 3.4). In the context of MFP, an extension to this methodology was proposed by

Dunkler et al (Dunkler et al, 2016). They suggested providing 'joint' shrinkage factors for semantically related regression coefficients, e.g., the components of a second-order FP. Systematic investigations of their proposal in the context of combined variable and functional form selection have not been reported.

### 5.2 Spline regression

"Traditional" selection techniques used in generalized linear models (GLMs) are also available for GAMs. Many R packages such as `mgcv` and `gamlss` implement approaches to forward or backwards stepwise regression based either on p-values of the smooth terms or on AIC/BIC type criteria. These approaches are intuitive and easy to use (and hence popular) but have the same weaknesses as GLMs. Additionally, since smoothness parameters are not fixed but estimated, p-values associated with the smooth terms are approximate and may be biased.

Alternatively, given a specific model structure, model selection might be performed using prediction error criteria or likelihood-based methods such as generalized cross-validation, Un-Biased Risk Estimator (UBRE), or REML with maximum likelihood for smoothness selection. Such methods, however, are unlikely to be successful and have known drawbacks (Wood, 2017).

More recent approaches consider penalization of smooth terms to perform variable selection. A penalty term on a regression spline penalizes the range space of the spline, controlling how wiggly the function is. For example, increasing the weight of a second-order derivative penalty to infinity will effectively force the term to linearity. On the other hand, it will not remove the term from the model, as the null space of the spline remains unpenalized. Marra and Wood (2011) therefore suggested null space penalization by adding an extra penalty for each smooth term to the log-likelihood. If all smoothing parameters for the term tend to infinity then it is penalized to zero and is effectively dropped from the model. There also exists a special type of shrinkage smoothers, based on cubic regression splines and thin plate regression spline smoothers in which a modification of the smoothing penalty would allow strong enough penalizations to shrink all coefficients of the smooth term to zero (Wood, 2017). Clearly, such penalization methods are not easily implemented but rely on the availability of a software package (such as `mgcv`) to provide the appropriate fitting routines as well as methods for optimization of the smoothing parameters. Clearly, since penalization techniques are relatively new, further research is needed to compare their performance with that of stepwise procedures in GAMs.

Other approaches include the Component Selection and Smoothing Operator (COSSO, Lin and Zhang, 2006), SpAM (Ravikumar, Liu, Lafferty and Wasserman, 2008), and the method of Meier et al. (Meier, van de Geer and Bühlmann, 2009), all based on a penalized likelihood that reduces the problem to Lasso-type estimation (Tibshirani, 1996). Chouldechova and Hastie (Chouldechova and Hastie, 2015) introduced Generalized Additive Model Selection (GAMSEL) that chooses between linear and non-linear fits for the component functions. Antoniadis et al (Antoniadis, Gijbels and Verhasselt, 2012) discuss variable selection using P-splines, based on an extension of the nonnegative garrotte. These methods have not been evaluated in practice and further research is needed to understand their properties.

We finally note that the variable selection procedure of the component-wise boosting algorithms described in Section 3.2.2 also applies to spline regression. In fact, component-wise boosting can be used to fit a large variety of GAMs, allowing a forward stagewise selection of nonlinear effects and an automated determination of the appropriate amount of smoothing (Tutz et al, 2006; Schmid et al, 2008; Hofner, Hothorn, Kneib and Schmid, 2011).

Royston and Sauerbrei proposed transferring the MFP approach to regression splines and smoothing splines (Royston and Sauerbrei, 2007; Royston et al, 2008). The MFP algorithm provides a principled approach for systematic selection of variables and FP functions. In the multivariable regression splines (MVRS) algorithm they combined backward elimination with a search for a suitable restricted cubic spline function. In the multivariable smoothing splines (MVSS) algorithm, they replace the FP components by cubic smoothing splines. For more details and a comparison of results from MFP, MVRS and MVSS we refer to Royston and Sauerbrei (Royston et al, 2008, chapter 9).

## 6. Examples illustrating the problems

In this section, we illustrate by way of selected published analyses of observational studies that guidance on many aspects of multivariable modelling outlined above is urgently needed.

### 6.1 A case of popular but highly problematic variable selection

Ramaiola et al (2015) conducted linear regression analyses to identify prognostic factors of plasma prolifin-1 (Pfn-1) levels in myocardial infarction patients with ST-segment elevation. Their study sample comprised 86 patients. Twelve explanatory variables were considered: duration of ischaemia from onset of pain to percutaneous coronary intervention, age, gender, diabetes, hypertension, obesity, dyslipidaemia, tobacco, antithrombic treatments, and levels of hsC-reactive protein and P-selectin. The authors first used bivariate correlation analysis or t-tests to assess the association between Pfn-1 and each of the potential explanatory variables. Variables with a significant ($\alpha = 0.05$) association with Pfn-1 were then taken as the starting set in a multivariable model to which backward elimination ($\alpha = 0.05$) was applied. Finally, two variables, duration of ischaemia and patient age 'remained as independent factors for Pfn-1 levels', both with highly significant p-values lower than 0.001. Surprisingly, time of ischaemia was not mentioned as a candidate variable in the statistical methods section. The values of the estimated regression coefficients were reported, but not their standard errors, and for interpretation purposes, no explicit reference was made to the measurement scale of the corresponding variables. No assessment of model assumptions, nonlinearities or interactions was mentioned. The stability of the model was neither investigated nor questioned.

### 6.2 Example of conflicting results for a variable with a spike at zero

Arem et al (2015) report on a pooled analysis of 666,137 participants of 6 studies from the National Cancer Institute Cohort Consortium, investigating the dose-response relationship between leisure time physical activity (LTPA) and mortality. LTPA is measured based on self-reported average hours per week spent on several physical activities over the prior year. These hours per week were then transformed into metabolic equivalent (MET) hours per week to adjust for different energy expended by type of activity. MET is a continuous variable with a spike at zero (8.0%). The main analysis adjusts MET for several covariates, e.g., sex, smoking, alcohol (categorized into 4 groups), educational level, marital status, history of cancer, history of heart disease, and body mass index (categorized into 5 groups). Age was used as the time scale in Cox regression analysis. While the supplement contains a figure in which MET entered the analysis using restricted cubic splines, the main paper provides results for categorized MET (7 categories including 0 as a category). Details of the spline specification are not given. Both analyses with 0 MET hours per week as the reference suggest a benefit from physical activity, but do not agree well numerically. For example, the hazard ratio for physical activity between 12 minutes and 7.5 MET hours per week is 0.8 (95% confidence interval 0.78 to 0.83), while in the same range of physical activity, the spline analysis suggests a hazard ratio smoothly changing from 1 to about 0.82 at 7.5 hours per week. Also, all other hazard ratios of the categorized analysis are more extreme than would be suggested by the spline analysis. The paper contains no explanation of the discrepancy. We suspect that the less extreme hazard ratios of the spline analysis could be caused by ignoring the spike at zero in that analysis. For patient counselling, the categorized analysis is misleading as it suggests a reduced hazard of mortality (hazard ratio 0.8) already at very low levels of MET (12 minutes per week), which seems implausible. This is even more crucial as the paper contains the misleading concluding statement 'These findings … provide important evidence to inactive individuals by showing that modest amounts of activity provide substantial benefit for postponing mortality'. Since the two analyses give different results, we must conclude that such a statement is not fully supported by the data, implying that at least one analysis is questionable.

## 7. Towards state of the art – research required!

In the earlier sections, we have illustrated that many variable selection procedures are available. Unfortunately, for all of them knowledge of their properties and the number of informative comparisons are limited. In all sections we raised many issues requiring further research. We identified several papers In the literature providing recommendations for practice (e.g., Harrell, Lee and Mark, 1996; Harrell, 2001; Harrell, 2015; Sauerbrei et al, 2007; Heinze et al, 2018) or, as expressed by Royston and Sauerbrei (2008), 'Towards recommendations …under the assumptions …'. They state that clear guidance is almost always impossible and stress that guidance can only be obtained under specific assumptions (e.g., large enough sample size, absence of interactions, etc). Based on the earlier sections, in Table 1 we summarize issues we consider as being the most relevant for further investigations. Theoretical results derived for specific 'sub-problems' and under strong assumptions may supply important insight, but in general we need empirical evidence based on extensive simulation studies. Difficulties in conducting unbiased and informative simulation studies were discussed in a letter from the STRATOS simulation panel (Boulesteix, Binder, Abrahamowicz and Sauerbrei, 2018). The need to conduct better designed and analyzed simulation studies has been expressed (Morris, White and Crowther, 2018).

Table 1 about here

*Issue 1 Investigation and comparison of the properties of variable selection strategies:* Traditional and modern variable selection approaches need to be compared in a head-to-head manner in a variety of explanatory modelling situations. To be valuable for explanatory modelling, target parameters should be those that influence the 'correct' inclusion and exclusion of variables, model stability, and model complexity. They should address the accuracy of the effects to be estimated, i.e., the regression coefficients. The number of candidate variables, their distributions, the correlation structure, the strength of effects and the sample size should be the main meta-parameters to be varied.

*Issue 2 Comparison of spline procedures in a univariable and multivariable context:* In the univariable situation the important questions are whether the results derived with the many suggested spline procedures differ (substantially) from the true function, and how much that depends on relevant parameters (e.g., number of knots). Furthermore, permitted complexity of the function and usability of procedures for non-experts are relevant. Criteria in the multivariable situation are similar, with an additional focus on correct variable inclusion and exclusion. We should consider modifying some of the procedures proposed in light of results from univariable investigations. The focus should lie on practical data modelling situations with multiple variables of mixed types, with splines used for modelling continuous variables. Different types of functions and sample sizes should be the main meta-parameters in the univariable situation. Additional parameters (as with Issue 1) are required in the multivariable situation.

*Issue 3 How to model one or more variables with a 'spike-at-zero'?* For one variable, a two-step FP-based procedure (see section 4.3) has been proposed. Using splines, a one-step approach would be possible and comparisons of various spline based approaches with the FP approach are needed. Main criteria are similar to those mentioned for the univariable situation in Issue 2. The FP approach has been extended for two spike-at-zero variables. Four strategies have been suggested, but current knowledge is limited.

*Issue 4 Comparison of multivariable procedures for model and function selection*: Given the huge number of penalized and spline-based approaches recently proposed (see section 5.2), additional studies such as Binder et al (Binder, Sauerbrei and Royston, 2013) are needed to evaluate their performance. The emphasis should be on accuracy, efficiency, transportability, ease of implementation and interpretability of the resulting multivariable models.

*Issue 5 Role of shrinkage to correct for bias introduced by data-dependent modelling:* The usefulness of post-selection shrinkage methods in removing overestimation bias in the regression coefficients and estimated functional forms in multivariable models needs further investigation. Guidance in selecting the right type of shrinkage factors (parameterwise, joint or global) and cross-validation procedure (jackknife, bootstrap) to obtain stable shrinkage factors is needed. Comparison with approaches combining variable selection and shrinkage (see section 3.3.1) are needed. Post-selection shrinkage can be generally used for models combining variable and functional form selection. This context needs to be investigated.

*Issue 6 Evaluation of new approaches for post-selection inference:* Recently, new procedures for obtaining confidence intervals for regression coefficients after applying forward selection, the Lasso or LARS have been proposed, but they have not yet been evaluated in typical explanatory modelling situations. For traditional variable selection methods, the performance of alternative variance estimators that are robust to model misspecification have not yet been investigated.

*Issue 7 Adaption of procedures for very large sample sizes needed?* So far, nothing has been said about building of interpretable explanatory models with large data sets. In recent years, more and more enormous data sets, e.g., with sample size n > 100,000 and number of potential explanatory variables p > 1,000 have become available for medical research, e.g., in pharmacoepidemiology, in studies with electronic health records or registries, and in individual patient data meta-analyses. Procedures for multivariable model building were developed with much smaller numbers in mind, with respect to both n and p. Approaches considering the natural or empirical grouping of variables ('variable clustering') may become more relevant here, while the usual settings for tuning parameters for variable selection procedures need to be rethought. Alternative approaches that optimally combine subject-matter knowledge with statistical learning need to be developed and evaluated.

Topics raised in Table 1 will keep our topic group (and hopefully others) busy for several years, but what is the alternative for developing evidence supported guidance for the selection of variables and functional forms in multivariable analysis?

8. Discussion

In a joint effort, members of TG2 identified seven issues which we consider key to building a multivariable model with continuous variables. The list could be extended, as other experts in the field may have different experiences and preferences and may consider other issues more important. We welcome discussion and critique as this would trigger urgently needed research in one of the most important areas of statistical analysis. We would also be pleased if other experts and research groups were motivated by section 7 and decided to do further research on issues we raised. Members of our group have considerable expertise to guide the necessary research we outlined. We consider this work an important step on the long journey towards evidence-supported guidance on the state-of-the-art. Following the general approach of STRATOS, guidance for experienced analysts will later be adapted to researchers at level 1, who have less statistical training.

It is well known that the aim of a study strongly affects the choice of a suitable analysis. Here we concentrate on models for description and mention that some things change when the main aim is prediction. That is one of the reasons why STRATOS has a separate topic group for prediction (TG6). We also ignore the recent trend in epidemiological studies to use directed acyclic graphs (DAG) for specifying and deriving confounders to enter regression models. It is a principled approach requiring critical decisions based on subject-matter knowledge. Nevertheless, it may be difficult to handle when there are many variables and it does not help in dealing with selection of functional forms for continuous variables. While in general we do not discourage thinking about a DAG to identify a meaningful working set of variables, the topic is perhaps more relevant for causal inference. Therefore, we leave it for STRATOS topic group TG7 ('Causal inference'). Naturally, there are many other questions we have ignored because they are covered by different topic groups. For example, each analysis is preceded by many decisions in preparing the data during an initial data analysis, and these decisions can have an important influence on the results. For a detailed discussion see the paper of topic group TG3 ('Initial data analysis') (20). Nearly all the issues we discuss apply for researchers working with survival data (TG8) or high-dimensional data (TG9).

It is difficult and still unusual to pre-specify a detailed analysis plan for observational studies, as is standard for the analysis of RCTs. Consequently, we must assume that usually several analyses are conducted, but that results are presented only for some of them. It is well-known that final models derived data-dependently are at least partly a result of chance. It is still common practice to base inference on a "conditional model", as if the model had been specified *a priori*. However, we must realize that because the uncertainty of the model-building process is ignored, the parameter estimates may be (heavily) biased and their variances are usually underestimated.

With the increasing power of computing environments, it became easier to use Bayesian approaches or resampling methods aimed at improving predictions and their variances by using model averaging approaches (Chatfield, 1995, Draper, 1995, Hoeting, Madigan, Raftery and Volinsky, 1999; Buckland, Burnham and Augustin, 1997; Augustin, Sauerbrei and Schumacher, 2015). Attempts to quantify model uncertainty have been known since the eighties (Chen et al, 1985; Altman et al, 1989). Researchers stress that data-dependent derivations of multivariable models need to be complemented by such investigations (Sauerbrei et al, 2015). Unfortunately, investigations of model stability are a long way from becoming standard practice.

Regarding issues raised in section 7, we emphasized that mathematical theory is unlikely to help and that appropriate simulation studies are the key tool to assess and compare the properties competing approaches. However, it is well known that many simulation studies comparing a suggested new method with existing methods may be (strongly) biased in favour of the new method (Boulesteix et al, 2018). Morris et al (2018) provided empirical evidence that simulation studies are often poorly designed, analysed and reported. They provided a structured approach to planning and reporting simulation studies, which involves defining aims, data-generating mechanisms, estimands, methods and performance measures ('ADEMP'). STRATOS has a simulation panel which has started to work on guidance for the design, conduct and reporting of simulation studies. The corresponding authors of the two papers just mentioned are members of this panel.

We outline the research needed to determine which procedures could be considered state-of-the-art (SOTA). To use SOTA methodology, an experienced scientist with deep knowledge of statistical methodology is required. However, most analyses are conducted by people with a lower level of statistical knowledge who need assistance working with the more basic and straightforward methods. Many analyses may be done with basic, simple methods not requiring advanced methods regarded as SOTA. One important aim of the STRATOS initiative is to formulate guidance for analysts with a lower level of statistical knowledge. See Sauerbrei et al (2014) for an outline of the proposed strategy.

During the last two decades, reporting guidelines have been developed for many types of studies, with the EQUATOR network acting as an umbrella. For some of these guidelines, so-called 'explanation and elaboration' papers have been published. They include statements and some advice on statistical analysis, but they discuss some more basic issues and aim more at analysts with a lower level of statistical knowledge. The STROBE, REMARK and TRIPOD guidelines are probably the most pertinent to issues discussed in this paper (Vandenbroucke et al, 2007; Altman, McShane, Sauerbrei and Taube, 2012; Moons et al, 2015). To avoid selective reporting, which gives rise to biased results and misleading interpretation, it is important to report all analyses, preferably by giving key information via a structured display. Altman et al (2012) proposee the REMARK profile as a suitable means of improving reporting of prognostic marker studies. Structured displays still need to be developed for other types of studies.

The overarching aim of the international STRengthening Analytical Thinking for Observational Studies (STRATOS) initiative is to provide accessible and accurate guidance documents for relevant topics in the design and analysis of observational studies. The topic group TG2 working on guidance for the selection of variables and functional forms has shown that many open issues exist and that much research is needed to fill the evidence gap when trying to determine which approaches may be termed 'state-of-the-art'.

## 9. Acknowledgment

This work was supported by the Deutsche Forschungsgemeinschaft [SA580/10-1] to Willi Sauerbrei, by the European Commission's programme Erasmus+ Staff Mobility for Training during the fellowship of Georg Heinze in Freiburg in November 2018 and by UK Medical Research Council programmes MC to Patrick Royston [UU_12023/21, MC_UU_12023/29]. We thank Tim Haeussler, Andreas Ott and Christine Wallisch for administrative assistance.

The international STRengthening Analytical Thinking for Observational Studies (STRATOS) initiative aims to provide accessible and accurate guidance for relevant topics in the design and analysis of observational studies (http://stratos-initiative.org).

Members of TG2 at the time of first submission: Michal Abrahamowicz, Heiko Becher, Harald Binder, Daniela Dunkler, Frank Harrell, Georg Heinze, Aris Perperoglou, Geraldine Rauch, Patrick Royston, Willi Sauerbrei, Matthias Schmid

**Tables**

| Table 1: Relevant issues in deriving evidence-supported state-of-the-art guidance for multivariable modeling ||
|---|---|
| 1 | Investigation and comparison of the properties of variable selection strategies |
| 2 | Comparison of spline procedures in both univariable and multivariable contexts. |
| 3 | How to model one or more variables with a 'spike-at-zero'? |
| 4 | Comparison of multivariable procedures for model and function selection |
| 5 | Role of shrinkage to correct for bias introduced by data-dependent modelling |
| 6 | Evaluation of new approaches for post-selection inference |
| 7 | Adaption of procedures for very large sample sizes needed? |

**Web supplement**

1. Methods based on spline functions

Splines (piecewise polynomial functions) come in many shapes and forms. A basic dichotomy is between regression splines, which focus on the polynomial choice and finding a set of knot locations, and smoothing splines, which focus on minimization of a penalty function.

Regression splines are particularly attractive due to their simplicity, ability to be included in any regression software and relatively simple mathematical form. Although there exist several different types of spline basis functions (including truncated polynomials, B-splines, cubic splines, natural splines), in practice the type of basis chosen should not dramatically alter the fitted function. The choice of basis depends on numerical stability and interpretability. Two preferred approaches are B-splines, which are numerically stable, and natural cubic splines, which are constrained to linearity in the tails.

With both B-splines and natural splines the number and placement of knots must be specified. Allowing a model to treat knot positions as parameters may increase flexibility but incur a heavy computational cost and instability. Automatic knot selection has also been suggested, typically based on stepwise procedures (Wand, 2000); however, this approach is computationally intensive and hard to implement in practice. Stone (Stone and Koo, 1985) has shown that provided there are enough knots to supply the required flexibility, the precise placement of knots is not crucial. Harrell (2017) suggests the use of four knots at appropriate quantiles for moderate samples sizes, three knots for smaller datasets (<30 observations) and five knots for larger datasets. Ruppert et al (Ruppert, Wand and Carroll,

2003) emphasize the benefit of unequally spaced knots, to avoid placement in empty regions of the domain of the continuous variable.

Smoothing splines extend cubic splines by placing a knot at each observation and adding a roughness penalty to control the smoothness of the fit. Eilers and Marx (1996) introduced P-splines by extending and simplifying ideas first seen in O'Sullivan (1986). Eilers and Marx's approach has been cited more than 3000 times in Google Scholar and has been implemented in most statistical software packages. P-splines utilize B-splines with a large number of equidistant knots which are controlled by a discrete penalty specified in terms of differencing operators. More recently, Eilers and Marx (2010), describing their experience of using P-splines, strongly recommended equally-spaced knots.

Thin-plate regression splines have some very desirable properties. Based on the work by Duchon (1977), Wood (2003) showed how to restrict thin-plate splines and provided a computationally efficient method. With Thin-plate regression splines, the analyst does not need to specify the number or placement of knots, nor to select a basis function. The method has become more widespread with the introduction of mgcv, a well-written and documented R package that allows for automatic penalty optimization using generalised cross validation, and an informative book on generalised additive models using R (Wood, 2006).

Although the mathematical properties of splines are well-understood, thorough comparison of methods is sparsely reported in the statistical literature. Ruppert et al (Ruppert, Wand and Carroll, 2009) described their experience of semiparametric regression in the years 2003-2007, Welham et al (Welham, Cullis, Kenward and Thompson, 2007) compared smoothing splines, P-splines and penalized splines using a truncated power function basis, and Binder et al compared splines with fractional polynomials. STRATOS recognizes the need for further in-depth review papers and systematic comparisons among methods. A review of spline procedures in R has been submitted for publication.

2. Fractional polynomials

With fractional polynomials (FPs), the aim is to extract full information from continuous variables in univariable and multivariable settings, resulting in models with simple and plausible functional forms. The selected model and functional forms should be interpretable from a subject-matter perspective. Interpretability, transportability and general usability of a model demand simplicity. Complex models, including complex functions of continuous covariates, are not useful when the aim is essentially descriptive.

A fractional polynomial is a simple function of a continuous covariate X. Starting from a straight-line model $\beta_1 X$, a natural extension is a power transformation model $\beta_1 X^p$. Royston and Altman (1994) formalized such a model by calling it a first-degree fractional polynomial or FP1 function, where the power p is chosen from a restricted set S ={−2,−1,−0.5, 0, 0.5, 1, 2, 3}, with 0 denoting log X. Extension of FP1 functions to the more complex and flexible two-term FP2 functions follows by defining FP2 functions with powers ($p_1$, $p_2$) as $\beta_1 X^{p_1} + \beta_2 X^{p_2}$, where $p_1$ and $p_2$ are taken from S. If $p_1 = p_2 = p$ the FP2 class is defined as $\beta_1 X_p + \beta_2 X_p \log X$, a so-called repeated-powers FP2 model. FP1 functions are monotonic and those with power p<0 have an asymptote as X→∞. FP2 functions may be monotonic or unimodal (i.e., have one maximum or one minimum for some positive values of X), and they have an asymptote as X→∞ when both $p_1$ and $p_2$ are negative. Generalization to FPm (m > 2) functions is straightforward, but the FP2 class is complex enough in many applications in the health sciences. The class of FP1 and FP2 functions is small (8 FP1 functions, 36 FP2 functions) but it includes a large variety of different functional forms. In contrast to splines, FPs are functions defined globally and cannot identify or respond to local features of the data to hand. For further details, see Royston and Sauerbrei (2008) or the website http://mfp.imbi.uni-freiburg.de/.

Function selection procedure

Royston et al (2008) defined a function selection procedure (FSP) based around a closed test procedure. The complexity of the finally chosen function is predicated on preliminary decisions as to the nominal p value ($\alpha$) and the degree (m) of the most complex FP model allowed. Typical choices are $\alpha$ =0.05 and m = 2 (FP2). With FP2 as the most complex allowed FP function, FSP selects an FP function according to the following procedure:

1. Test the best-fitting FP2 model for X at significance level $\alpha$ against the null model using 4 d.f. If the test is not significant, stop, concluding that the effect of X is "not significant" at the $\alpha$ level. Otherwise continue.

2. Test the best-fitting FP2 for X against a straight line at the $\alpha$ level using 3 d.f. If the test is not significant, stop, the final model being a straight line. Otherwise continue.

3. Test the best-fitting FP2 for X against the best FP1 for X at the $\alpha$ level using 2 d.f. If the test is not significant, the final model is FP1, otherwise the final model is FP2. End of procedure.

The test at step 1 is of overall association of the outcome with X. The test at step 2 examines the evidence for nonlinearity. The test at step 3 chooses between a simpler or more complex nonlinear model.

**References for Web supplement**